\documentclass[aps,prl,twocolumn,groupedaddress,showkeys]{revtex4}
\usepackage{epsfig}

\begin{document}


\title{Low concentrated hydroxyectoine solutions in presence of DPPC lipid bilayers: a computer simulation study}

\author{Jens Smiatek$^{1,2}$}
\email{smiatek@icp.uni-stuttgart.de}
\author{Rakesh Kumar Harishchandra$^3$}
\author{Hans-Joachim Galla$^3$}
\author{Andreas Heuer$^2$}
\affiliation{$^1$Institute for Computational Physics, University of Stuttgart, D-70569 Stuttgart, Germany\\
             $^2$Institute of Physical Chemistry, University of Muenster, D-48149 M{\"u}nster, Germany\\
             $^3$Institute of Biochemistry, University of Muenster, D-48149 M{\"u}nster, Germany}
\begin{abstract}
The influence of hydroxyectoine on the properties of the aqueous solution in presence of DPPC lipid bilayers is studied via semi-isotropic constant pressure (NPT) Molecular Dynamics simulations. 
We investigate the solvent-co-solute behavior in terms of Kirkwood-Buff integrals as well as hydrogen bond life times for an increasing hydroxyectoine concentration up to 0.148 mol/L. The observed preferential exclusion 
mechanism identifies hydroxyectoine as a kosmotropic osmolyte. 
Our findings in regards to the DPPC lipid bilayer indicate an increase of the surface pressure as well as the solvent accessible surface area in presence of higher hydroxyectoine concentrations. 
The results are in agreement to the outcome of recent experiments. With this study, we are able to 
validate the visibility of co-solute-solute-solvent effects for low and physiologically relevant osmolyte concentrations.
\end{abstract}
\maketitle



\section{Introduction}
Osmolytes allow extremophilic microorganisms to resist harsh living conditions \cite{Lentzen06,Driller08}. Typical examples for these species are ectoine and hydroxyectoine
which are zwitterionic, strong water binding and low-molecular weight organic molecules. 
The functionalities of these molecules in living organisms among others are given by the protection of protein conformations \cite{Lentzen06,Driller08,Yu04,Yu07,Smiatek12} and the 
fluidization of lipid membranes \cite{Galla10,Galla11}. The protective properties become mainly important under environmental stress conditions, e. g. high temperature, extreme dryness and salinity \cite{Knapp99}.
Several studies have identified that several combinations of osmolytes can be found in biological cells. The concentration of a single osmolyte in these mixtures varies between 0.1 to 1.0 mol/L (\cite{Yancey2005} and references
therein).  
Due to the fact that a lot of osmolytes are not affecting the cell metabolism, specific molecules like the ectoines are also commonly called compatible solutes.\\
Recent studies were focusing on the
molecular functionality of the ectoines as well as the analysis of the protective behavior \cite{Yu04,Yu07}. 
The theoretical framework for the explanation of co-solute-solute effects has been mainly established in terms of the preferential exclusion \cite{Timasheff02} and the transfer free energy model \cite{Rose2008}.\\
These models focus on the strong ordering of the local water shell around the osmolytes and the exclusion from the immediate hydration shell of the solute which 
results in a preferential hydration behavior and a stabilization of the solutes native structure \cite{Timasheff02,Rose2008}. 
Among these, more refined versions of theories have been in addition
published which explicitly rely on the properties of chaotropic and kosmotropic behavior \cite{Collins97,Collins04,Collins07,Ninham12} and the corresponding interaction with macromolecular 
surfaces. Co-solutes which strengthen the water hydrogen network are called kosmotropes (structure makers) while osmolytes which weaken the water network structure are 
called chaotropes (structure breaker). The separation of co-solutes into these two species is not unique and straightforward \cite{Ninham12}. 
Interactions and binding properties between kosmotropes and chaotropes can be predicted by the 'law of matching water affinities' \cite{Collins04,Ninham12}.
A main point of this theory is the investigation of the corresponding hydration free energies which loosely depend among other factors on the molecular charge \cite{Ninham12}. 
One of the major achievements of the 'law of matching water affinities' is the molecular description of a repulsive behavior for kosmotropic osmolytes from polar surfaces and vice versa, 
the attraction of chaotropic agents like urea. 
The preferential binding of urea has been validated in recent computer simulations \cite{Horinek11} while additional studies have observed a kosmotropic behavior for ectoine in terms of a preferential exclusion mechanism
around Chymotrypsin Inhibitor II \cite{Yu07}.\\
It has been stated that kosmotropic co-solutes typically accumulate in the second or third hydration shell of the solvated macromolecule.
In regards to their high charging and affinity for water molecules, it is assumed that this appearance strongly influences the first and the second hydration shell of the polar solute. 
The consequence of this behavior is given by a diminished 
number of solute-water hydrogen bonds which is compensated by a significant shrinkage of the solute surface to maintain 
a constant hydrogen bond surface density. It is commonly believed that this shrinkage in size is the molecular reason for the preservation of native protein conformations 
in presence of kosmotropic co-solutes \cite{Collins04,Ninham12}.\\
Although the stabilizing effects on proteins in presence of specific osmolytes have been studied extensively before, 
less is known about compatible solutes and their interactions with bilayers. A small number of theoretical studies have focused on
sugars like trehalose and their interactions with lipid membranes and monolayers \cite{dePablo03,Hunenberger04,Pastor05,Hunenberger06,Sum06,Hunenberger08,Hunenberger10}.
For high molar concentrations of trehalose, it has been found that replacement of water molecules by the formation of additional sugar-membrane
hydrogen bonds plays a major role \cite{Hunenberger04}. Despite this
interpretation, it has been also discussed that
the effects observed in sugar-DPPC mixtures can be only systematically explained by an interplay of several 
mechanisms \cite{Hunenberger10}.\\ 
In addition to theoretical studies, experimental findings have indicated a significant broadening of the liquid expanded (LE) - liquid condensed (LC) 
phase transition of monolayers in presence of ectoine and hydroxyectoine \cite{Galla10,Galla11}.
This was mainly indicated by the study of the corresponding 
surface pressure area isotherms.  A main result of these studies was the observation of a surface pressure increase for higher hydroxyectoine
concentrations.  
In addition it was supposed that the domain sizes of the liquid condensed regions
significantly shrink in presence of hydroxyectoine which corresponds to a variation of the line tension \cite{Galla10,Schwille07,Baumgart08,Ruckenstein97}. 
In regrds to the biological function, the above mentioned effects 
are of particular important for signaling processes and cell repair \cite{Driller08,Galla10}.\\ 
In regards to the preferential exclusion/binding behavior of osmolyte-solute-solvent mixtures, computer simulations allow a detailed study of the corresponding molecular mechanisms.
A theoretical framework which allows to distinguish between exclusion and binding behavior has been established in terms of the Kirkwood-Buff theory of solutions \cite{Kirkwood51,Ben-Naim92,Trout03}. The corresponding
analysis has been therefore successfully applied to the study of urea and polyglycine interactions \cite{Horinek11}. 
It has been shown that the calculation of the Kirkwood-Buff integrals allows the effective determination of transfer free energies in addition to the detection of kosmotropic as well as 
chaotropic behavior \cite{Yu07,Horinek11}.\\
In this paper, we study the properties of an aqueous hydroxyectoine solution in presence of DPPC lipid bilayers via semi-isotropic constant pressure (NPT) all-atom Molecular Dynamics simulations.
The concentration of hydroxyectoine is low but physiologically relevant \cite{Yancey2005} with a maximum value of 0.148 mol/L. We have been inspired to use these small concentrations due to recent experimental 
findings for aqueous hydroxyectoine-DPPC monolayer mixtures \cite{Galla10,Galla11}. 
Most of the simulation studies usually employ high co-solute concentrations which are often above one mole per liter to study pronounced behavior at
unphysiological conditions \cite{Yu04,Yu07,Horinek11}. 
With this study, we are able to validate the observation of effects at smaller concentrations in agreement to experimental findings.\\
Our main results include the characterization of hydroxyectoine as a kosmotropic osmolyte which strengthens the water hydrogen bond network. 
We are further able to validate a weakening of DPPC-water hydrogen bond interactions in presence of hydroxyectoine. 
In regards to the DPPC lipid bilayer properties, our results validate an increase of the surface pressure and the solvent-accessible surface area in agreement to the experimental 
results \cite{Galla10,Galla11,Smiatek12}. We emphasize the importance of electrostatic interactions between co-solutes and solutes
for the understanding of the observed effects by the calculation of the bilayer electrostatic potential.\\
The paper is organized as follows. In the next section we shortly introduce the theoretical background. In the third section we illustrate the simulation details and the methodology. 
The results for the solvent properties and the DPPC lipid bilayer are presented and discussed in the fourth section.
We briefly conclude and summarize in the last section. 
\section{Theoretical Background}
\subsection{Kirkwood-Buff integrals and preferential binding parameter}
The evaluation of statistical mechanics methods on the co-solvent and solvent distribution function allows important insights into the preferential exclusion as  
well as binding behavior in terms of the corresponding Kirkwood-Buff theory which has been introduced in the early 1950's \cite{Kirkwood51,Ben-Naim92}.
The radial distribution function of molecules or atoms $\beta$ around solutes $\alpha$ can be expressed by
\begin{equation}
\label{eq:rdf}
g_{\alpha\beta}(r) = \frac{\rho_{\beta}(r)}{\rho_{\beta,\infty}}
\end{equation}
where $\rho_{\beta}(r)$ denotes the local density of $\beta$ at a distance $r$ around the solute and $\rho_{\beta,\infty}$ the global density in the bulk phase \cite{Leach01}.
The Kirkwood-Buff integral is given by the integration of Eqn.~\ref{eq:rdf} 
\begin{equation}
\label{eq:KBI}
G_{\alpha\beta} = \lim_{R\rightarrow\infty} G_{\alpha\beta}(R) = \lim_{R\rightarrow\infty}\int_{r=0}^{r=R}4\pi r^2(g_{\alpha\beta}(r)-1)dr
\end{equation}
where the above relation is valid in the limit of $R=\infty$ \cite{Yu07,Horinek11,Kirkwood51,Ben-Naim92,Trout03}.
Eqn.~\ref{eq:KBI} can be used to calculate the excess coordination number of molecules or atoms of $\beta$ (hydroxyectoine) around $\alpha$ (DPPC) via \cite{Yu07,Kirkwood51,Ben-Naim92}
\begin{equation}
N_{\beta}^{xs} = \rho_{\beta,\infty}G_{\alpha\beta}= \rho_{\beta,\infty}\lim_{R\rightarrow\infty} G_{\alpha\beta}(R)
\end{equation}
which allows to evaluate the preferential binding coefficient $\nu_{\beta\gamma}(R)$ under the assumption of finite distances $R$ 
with
\begin{equation}
\label{eq:bind}
\nu_{\beta\gamma}(R) = \rho_{\beta,\infty}(G_{\alpha\beta}(R)-G_{\alpha\gamma}(R)) = N_{\beta}^{xs}(R) - \frac{\rho_{\beta,\infty}}{\rho_{\gamma,\infty}}N_{\gamma}^{xs}(R)
\end{equation}
where the indices $\gamma$ represent solvent molecules while $\alpha$ and $\beta$ are the solute DPPC and the co-solute hydroxyectoine.
A negative value for the preferential binding coefficient of Eqn.~\ref{eq:bind} implies a preferential exclusion of hydroxyectoine from the lipid bilayer surface while a positive value indicates a
preferential binding \cite{Yu07,Trout03}. 
\subsection{Hydrogen bonds and  water relaxation times}
The formation of lipid bilayers in aqueous solutions is mainly driven by the hydrophobic effect \cite{Ball08}. The detailed molecular mechanism for the hydrophobic effect is still heavily
under debate but it is consensus that the formation and cleavage of hydrogen bonds is of main importance \cite{Ball08}.
For a detailed investigation of the co-solute-solvent properties, we have analyzed the water hydrogen bond characteristics in presence 
of varying hydroxyectoine concentrations.\\
We apply the Luzar-Chandler definition of hydrogen bonds \cite{Luzar96,Luzar2000}, which restricts a maximum length of 0.35 nm between the interacting oxygen and hydrogen atom and an angle of not more than 30 degrees.
The value for the life time 
allows an estimate of the relative strength for the corresponding hydrogen bond. This argument is inspired by results of transition state theory 
which connects the rate constant $k_F$ to the activation free energy $\Delta G^{*}$ via
\begin{equation}
\label{eq:rate}
k_F = \frac{1}{\tau_F}= \frac{k_BT}{h} \exp\left(-\frac{\Delta G^{*}}{k_BT}\right)
\end{equation}
with the thermal energy $k_BT$, the forward life time $\tau_F$ and the Planck constant $h$ \cite{vanderspoel06}. 
In terms of the underlying statistical analysis of hydrogen bonds \cite{vanderspoel06},
Eqn.~\ref{eq:rate} reveals that longer life times correspond to larger free activation energies which indicates a strengthening of the hydrogen
bond network.\\
In addition to the hydrogen bond analysis, we have also calculated the dipolar reorientation time of water molecules in presence of hydroxyectoine. The evaluation of this quantity gives access to an estimation 
of the translational solvent entropy. It has been discussed that solvent entropies may play a significant role in contributions to the solvation free energies and therefore the exposure of hydrophilic and 
hydrophobic surfaces \cite{Ball08,Finkelstein,Ninham12}.
To study this property, we have investigated the autocorrelation time for the dipolar orientation $\vec{\mu}$ between two water molecules in presence of DPPC lipid bilayers and hydroxyectoine which is given by 
\begin{equation}
\label{eq:mu}
<\vec{\mu}(t)\vec{\mu}(t_0)>\sim\exp(-t/\tau)^\beta.
\end{equation}
It can be seen that the autocorrelation function follows a stretched exponential behavior which is dependent on the exponent $\beta$ \cite{Stanley05}. The evaluation of the corresponding times $\tau$ 
allows a quantitative determination of the influence of hydroxyectoine on the water dynamics. 
\subsection{Surface pressure}
As it has been found out in recent experiments for DPPC monolayers in presence of hydroxyectoine \cite{Smiatek12,Galla10,Galla11}, a 
significant variation of the surface tension can be observed for increasing physiological hydroxyectoine concentrations. 
The surface tension can be calculated by 
\begin{equation}
\gamma = \int (P_N-P_T(z)) dz
\end{equation}
which can be also expressed for the ease of computation by 
\begin{equation}
\label{eq:gamma}
\gamma = \frac{1}{2}\left<L_z \left(P_{zz}-\frac{1}{2}(P_{xx}+P_{yy})\right)\right>
\end{equation}
where $P_N$ denotes the normal pressure with $P_{zz}$ as the z-component of
the pressure tensor and $P_T$ the transversal pressure given by
$1/2(P_{xx}+P_{yy})$ as defined by the x- and the y-components. The factor 
$1/2$ accounts for two interfaces in contact with water \cite{Pastor95,Kindt04,Tieleman07,Tieleman10}. The
box length in z-direction is denoted by $L_{z}$.\\
The surface pressure is given by
\begin{equation}
\label{eq:pi}
\Pi = \gamma_0-\gamma
\end{equation}
where $\gamma_0$ expresses the experimental surface tension of water at 300 K (71.6 mN/m) \cite{CPC} which allows a direct comparison with experiments as discussed in Ref.~\cite{Pastor05}.\\
\section{Simulation Details}
We have performed Molecular Dynamics simulations in explicit SPC/E water \cite{Straatsma87} with the software package GROMACS \cite{Berendsen95,Hess08,Spoel05}.
\begin{figure}[h!]
 \includegraphics[scale=0.35]{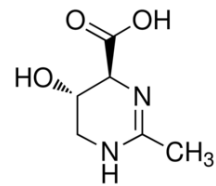}
\caption{Structure of neutral hydroxyectoine.}
\label{fig0}
\end{figure}
The chemical structure of hydroxyectoine ((4S,5S)-2-methyl-5-hydroxy-1,4,5,6-tetrahydropyrimidine-4-carboxylic acid) in its neutral form is presented in Fig.~\ref{fig0}. 
The derivation of the force field and the topology of hydroxyectoine is in detail described in Ref.~\cite{Smiatek12} where it has been also found that the zwitterionic form in aqueous solution is more stable than the neutral 
counterpart. We follow this finding by using purely zwitterionic molecules in our MD simulations.
The force field for the lipids and the starting structure with 64 DPPC molecules \cite{Tieleman} were modeled with the parameters presented in Ref.~\cite{Berger97}.\\
The Molecular Dynamics simulations have been carried out with periodic
boundary conditions. The simulation box has initial dimensions of
$(4.72450\times 4.23190\times 9.95050)$ nm$^3$. 
We performed simulations with $2,4,6$ and 8 hydroxyectoine molecules which correspond to effective concentrations of $0.037, 0.074, 0.111$ and $0.148$ mol/L. 
Electrostatic interactions have been calculated by the Particle Mesh Ewald sum \cite{Pedersen95}.
The time step was $\delta t=2$ fs and the temperature was kept constant by a Nose-Hoover thermostat \cite{Frenkel96}
at 300 K.
All bonds have been constrained by the LINCS
algorithm \cite{Fraaije97}.\\
After energy minimization and a 10 ns constant volume simulation to ensure the conformational equilibration of the lipid molecules \cite{Hunenberger10}, 
we conducted a 20 ns equilibration run followed by a 30 ns semi-isotropic constant pressure (NPT) data production simulation.
A Parrinello-Raman barostat has been used with a rescaling time step of 2 ps. The reference pressure in the x/y- and the z-direction was 1 bar and a compressibility of $4.5\times 10^{-5}$ bar$^{-1}$ was used. 
The solvent accessible surface area $\Sigma_{tot}$ was calculated by the sum of spheres centered at the
atoms of the studied molecule, such that a spherical solvent molecule can be placed in closest distance and in agreement to van-der-Waals interactions by following the constraint that other atoms are not
penetrated \cite{Scharf95}.\\
Hydrogen bonds have been defined as present if
the distance between the interacting atoms is less than 0.35 nm and the interaction angle is not larger than 30 degrees. The hydrogen bond density which is calculated for the DPPC lipid bilayers is given by the number of hydrogen
bonds divided by the hydrophilic solvent accessible surface area
$\rho_{HB} = <N_{HB}/\sigma_{HPL}>$.
\section{Results}
\subsection{Solvent properties}
\begin{figure}[tbh]
  \includegraphics[scale=0.35]{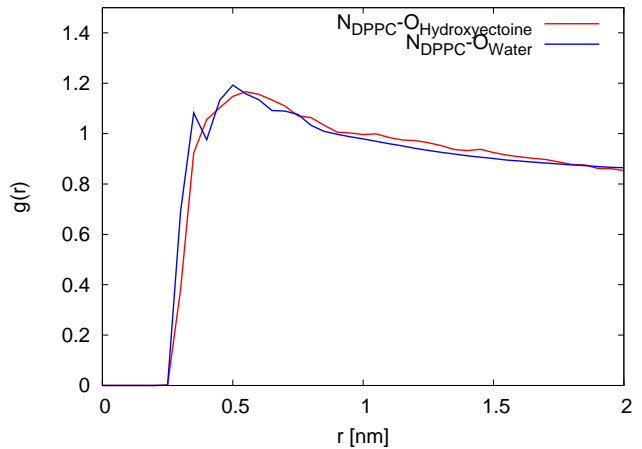}
  \caption{Pair radial distribution function $g(r)$ for nitrogen in DPPC ($N_{DPPC}$) and oxygen in hydroxyectoine (red line), respectively oxygen in water (blue line) for a hydroxyectoine concentration of 0.148 mol/L. 
           \label{fig1}}
\end{figure}

\begin{figure}[tbh]
  \includegraphics[scale=0.35]{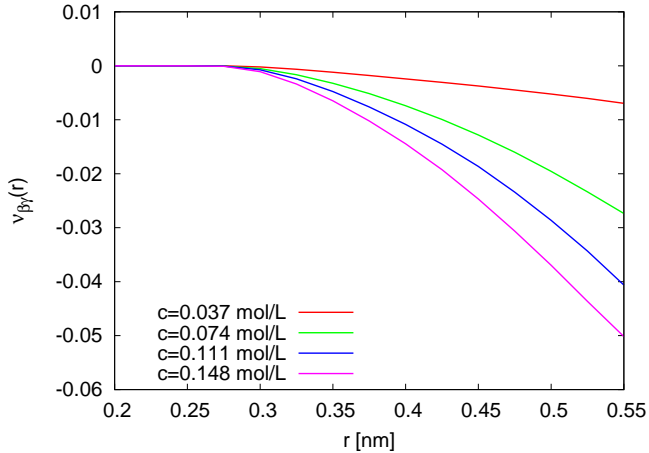}
  \caption{Preferential binding coefficient $\nu_{\beta\gamma}(R)$ for distances $r$ up to 0.55 nm for all four concentrations in presence of hydroxyectoine.}
           \label{fig2}
\end{figure}

\begin{figure}[tbh]
  \includegraphics[scale=0.45]{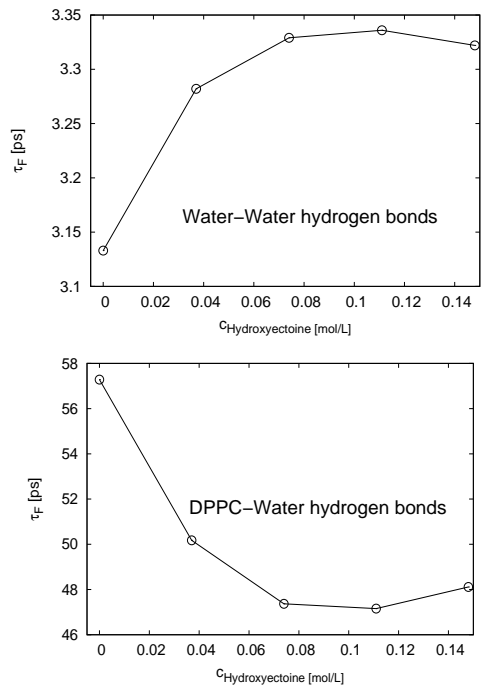}
  \caption{Water-water hydrogen bond forward lifetimes $\tau_F$ (top) and DPPC-water hydrogen bond life times (bottom) for increasing hydroxyectoine concentrations.}
           \label{fig3}
\end{figure}

The average position and the distribution of hydroxyectoine in front of the DPPC lipid bilayer can be easily determined by the evaluation of the pair radial distribution function (rdf). 
Due to the fact, that the nitrogen of DPPC is slightly positively charged ($+0.55 e$) and the oxygens in the carboxy group of hydroxyectoine are negatively charged ($-0.87 e$), we have decided to 
evaluate the rdf between these two atoms due to favorable electrostatic interactions. Furthermore we have calculated for a comparisoon the rdf for nitrogen in DPPC and oxygen in water molecules. 
The results for a hydroxyectoine concentration of 0.148 mol/L are shown in Fig.~\ref{fig1}.  
It can be seen that the pair radial distribution function $g(r)$ between DPPC and hydroxyectoine reveals a
significant appearance of the compatible solute at a distance of 0.35 to 0.85 nm
which is in good agreement to previous assumptions \cite{Smiatek12}. 
Comparing the pair radial distribution
function for the nitrogen and water oxygen to determine the position of the first hydration shell reveals that a large amount of
compatible solutes are accumulated at the second hydration shell of the
lipid bilayer. This can be validated by the smaller occurrence of
hydroxyectoine at 0.3 nm which roughly corresponds to the peak of the
first hydration shell.   
Thus it can be concluded that direct interactions between hydroxyectoine and DPPC in terms of hydrogen bonds are less important for these concentrations as it was also stated in \cite{Galla11}.\\ 
To further investigate the hydration behavior of DPPC, we have calculated the water hydrogen bond density at the DPPC hydrophilic solvent accessible surface area $\sigma_{HL}$. 
We have found a nearly constant value of $\rho_{HB}=7.43 \pm 0.01$ nm$^{-2}$ for the hydrogen bond density averaged over all hydroxyectoine concentrations. The results show no significant deviations concerning higher hydroxyectoine 
concentrations such that it can be assumed, that the concentration of the co-solutes does not affect the overall water hydrogen bond density at the DPPC surface. In addition, we have evaluated the number of direct 
contacts between hydroxyectoine and DPPC in terms of hydrogen bonds where we have found a value of $<N_{{HB}}>=0.6\pm 0.5$ for the highest hydroxyectoine concentration. All other values for lower concentrations are vanishing 
or nearly identical. Compared to the number of hydrogen bonds between water and DPPC and their 
contributions to the total energy, it can be concluded that the influence of the 
DPPC-hydroxyectoine hydrogen bonds is nearly negligible.
Regarding the water-replacement theory and compared to the 
results of trehalose-DPPC systems \cite{Hunenberger04}, it can be concluded that the reduction of the number of
water molecules in front of the DPPC lipid bilayer by hydroxyectoine as it has been assumed for proteins \cite{Collins04} is a minor effect which has not been detected in 
our simulations. Despite the additionally proposed diminished solvent accessible surface area \cite{Collins04}, instead we have observed an increasing surface area for the lipid bilayer. 
We will discuss this point in more detail in the next section. \\
For a detailed investigation of the preferential exclusion behavior, we have calculated the corresponding Kirkwood-Buff integrals \cite{Horinek11,Kirkwood51,Ben-Naim92,Trout03}.
The results for the preferential binding parameter (Eqn.~\ref{eq:bind}) are shown in Fig.~\ref{fig2}.
It can be clearly seen that the preferential binding coefficient $\nu_{\beta\gamma}(r)$ at lipid bilayer distances up to 0.55 nm is negative for all hydroxyectoine concentrations. 
The amount of exclusion increases with the concentration of the compatible solutes. This also
clearly states that hydroxyectoine is preferentially excluded from the DPPC bilayer surface due to energetic reasons \cite{Trout03}. 
Hence, the observed behavior is in good agreement to the predicted behavior for kosmotropic osmolytes \cite{Collins04}.\\ 
In the framework of recent theories, it has been stated that highly charged ions can be interpreted as kosmotropes (structure maker) \cite{Collins04}. 
These molecules tend to strengthen the water-water structure by increasing hydrogen bond lifetimes. 
The influence of chaotropic solutes like urea on the dynamics of water-water hydrogen bonds has been investigated in Ref.~\cite{Horinek11}. It was found that urea in contrast to kosmotropes decrease the strength 
of the water hydrogen bond network.
We have analyzed the corresponding characteristics for hydroxyectoine in terms of the hydrogen bond transition state theory and the orresponding statistical analysis \cite{vanderspoel06}. 
The corresponding forward life times of hydrogen bonds for water-water interactions and water-DPPC hydrogen bond interactions are shown in 
Fig~\ref{fig3}.
It can be clearly seen that hydroxyectoine leads to an increase of the hydrogen bond life times in presence of higher concentrations. Due to the fact that the forward life times are proportional 
to $\log(\tau_F)\sim \Delta G^*$, an increase of the life times also indicates a higher activation free energy barrier $\Delta G^{*}$. Hence the kosmotropic properties of hydroxyectoine are obvious and can be identified 
by a strengthening
of the water hydrogen bond network.
In addition, we 
have observed a saturation plateau of life times for hydroxyectoine concentrations $c\geq 0.074$ mol/L. The reverse behavior can be also observed for DPPC-water hydrogen bonds which means decreasing life times 
for increasing hydroxyectoine concentrations until a saturation plateau is reached. Thus, it can be concluded that the hydration properties of DPPC bilayers in terms of energetic contributions are slightly disturbed in 
presence of low molar hydroxyectoine solutions. 
These findings are in good agreement to the 'law of matching water affinities' which states that kosmotropic agents weaken the hydration behavior of polar solute surfaces \cite{Collins04}.  
Nevertheless, the pronounced influence of the DPPC lipid bilayer on the water dynamics can be also
observed in terms of long lifetimes as it was also discussed in a recent publication \cite{Debnath2010}. It was mentioned that in close
vicinity to the bilayer, a significant decrease of water diffusion coefficients can be validated.\\
Finally we have calculated the he hydrogen bond life times for hydroxyectoine and water where we have found a nearly constant values of $<\tau_F> = 4.27 \pm 0.22$ ps for all concentrations.  
\begin{figure}[tbh]
  \includegraphics[scale=0.35]{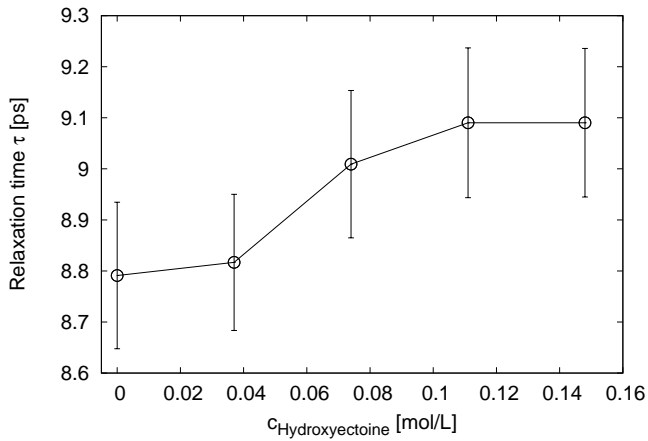}
  \caption{Increase for the dipolar autocorrelation time in presence of hydroxyectoine.}
           \label{fig4}
\end{figure}
To illustrate the importance of hydroxyectoine on the water dynamics and therefore the entropic contributions in terms of hydration behavior, we have also analyzed the water dipole reorientation times (Eqn.~\ref{eq:mu}). 
It has been stated that the general entropy of the water is significantly influenced by the number of intermolecular hydrogen bonds and the values for the reorientation times \cite{Ball08,Finkelstein}.
By the evaluation of the water dipole autocorrelation function for each concentration, we have identified a value for $\beta$ in Eqn.~\ref{eq:mu} of $0.86\pm 0.05$ which validates a stretched exponential behavior 
in agreement to recent results 
and theories \cite{Stanley05}. 
The corresponding relaxation times $\tau$ for each concentration are presented in Fig.~\ref{fig4}. We clearly observe an increase of the relaxation times in presence of hydroxyectoine which can be related to diminished entropic
contributions. The direct connection between the entropy and the relaxation dynamics has been discussed in Ref.~\cite{Bencivenga09} in which it has been shown that the relation $\tau = \tau_0\exp(-S/k_B)$ with the entropy $S$ 
is valid for several measurable relaxation times. 
Furthermore a plateau value for hydroxyectoine concentrations of $c\geq 0.074$ mol/L in agreement to the results of Fig.~\ref{fig3} can be also observed. Thus, a strong water structure influence effect can be observed in 
presence of hydroxyectoine.\\
To summarize the results of this subsection, we have validated that hydroxyectoine is repelled from DPPC lipid bilayer surfaces in terms of a preferential exclusion behavior. This effect is in good agreement
to recent theories concerning kosmotropic behavior for osmolytes. In a recent study \cite{Smiatek12} we were able to indicate a net accumulation of roughly 8-9 water molecules 
due to electrostatic interactions around hydroxyectoine. Combined with these results, the analysis of the hydrogen bond life times and the dipole reorientation times clearly reveals the kosmotropic behavior 
of hydroxyectoine. It has to be noted that the dynamic properties have been calculated by considering the complete number of water molecules, regardless if they interact with hydroxyectoine, DPPC or with themselves.
We are therefore confident that the change of global water dynamics in presence of low hydroxyectoine concentrations is a significant effect. 
\subsection{DPPC lipid bilayer properties}
\begin{figure}[tbh]
  \includegraphics[scale=0.35]{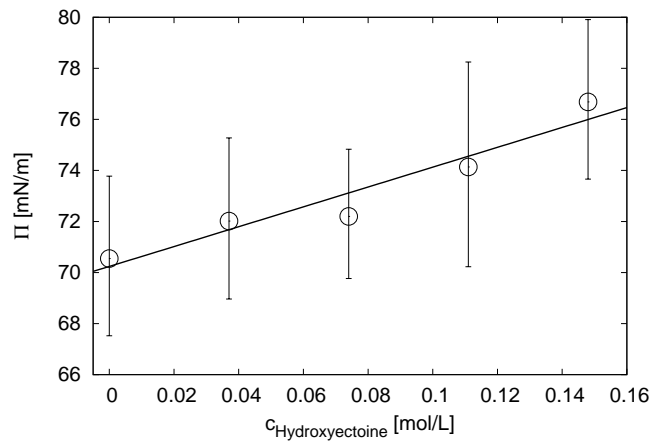}
  \caption{Surface pressure calculated by the relation $\Pi = \gamma_0-\gamma$
    with $\gamma_0 = 71.6$ mN/m. The presence of hydroxyectoine leads to a
    pressure difference of 6 mN/m for the highest concentration.} 
           \label{fig5}
\end{figure}

\begin{figure}[tbh]
  \includegraphics[scale=0.35]{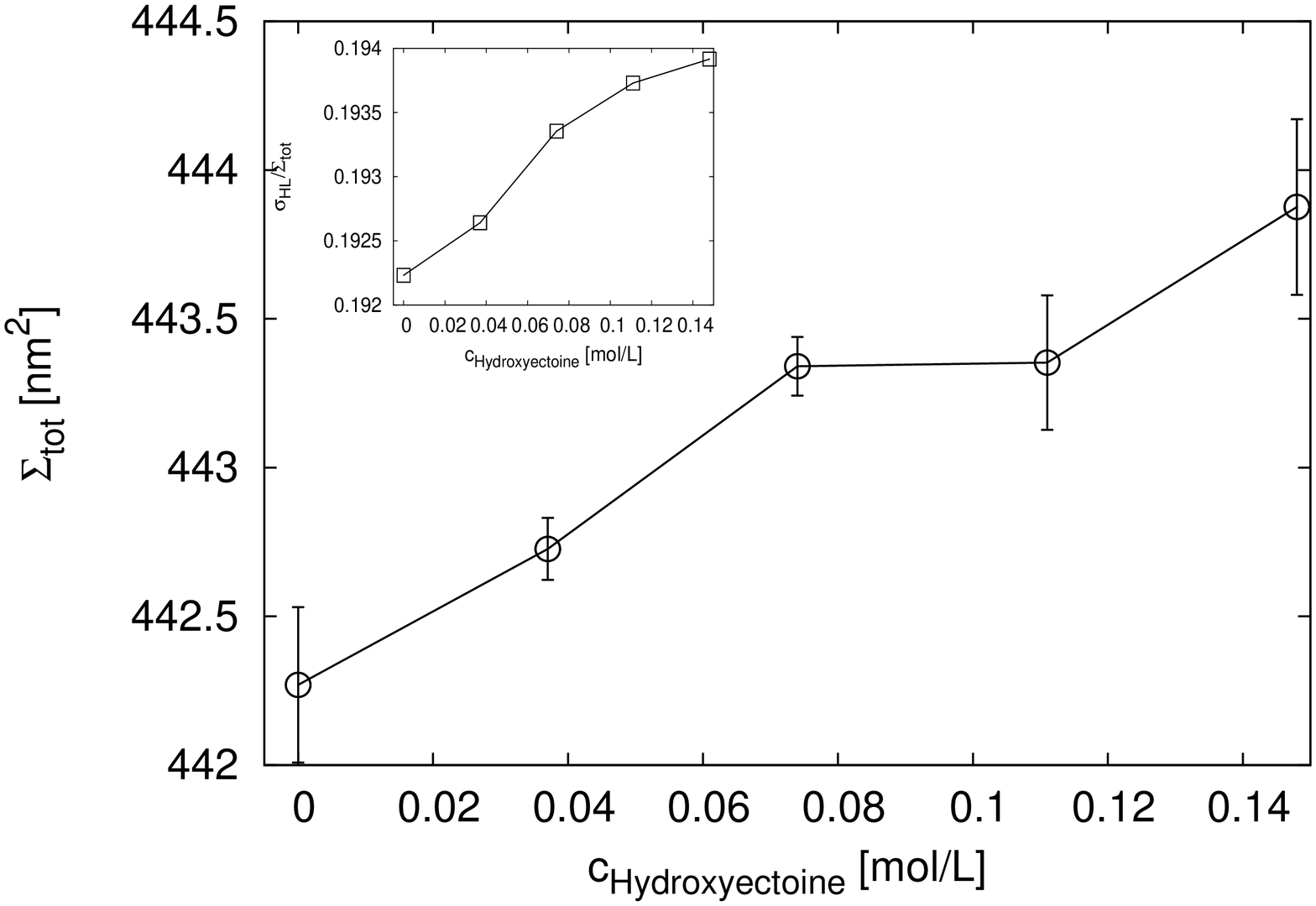}
  \caption{Total solvent accessible surface area for the DPPC lipid bilayer in presence of an increasing hydroxyectoine concentration. {\em Inset:} Ratio of the hydrophilic $\sigma_{HL}$ to the total solvent accessible surface 
           area $\Sigma_{tot}$.}
           \label{fig6}
\end{figure}

\begin{figure}[tbh]
  \includegraphics[scale=0.35]{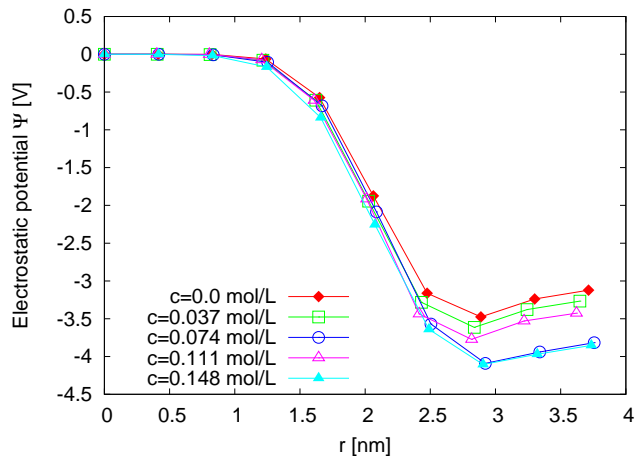}
  \caption{Electrostatic potential $\Psi$ for the DPPC lipid bilayer in presence of varying hydroxyectoine concentrations.}
           \label{fig7}
\end{figure}

In recent experiments for DPPC monolayers in presence of an aqueous hydroxyectoine 
concentration \cite{Galla10,Galla11,Smiatek12}, a broadening of the liquid expanded (LE) - liquid condensed (LC) phase transition was observed. This finding for hydroxyectoine concentrations around 0.1 mol/L 
has been validated in terms of surface pressure-area diagrams and has been compared to urea solutions where this behavior was found to be absent.\\
In order to prove the experimental results, we have calculated the surface pressure for varying hydroxyectoine concentrations according to Eqns.~\ref{eq:gamma} and \ref{eq:pi} for concentrations that are comparable 
to the experiments. 
The results for the surface pressure in presence of hydroxyectoine
are shown in Fig.~\ref{fig5}.
It becomes obvious that the presence of the osmolyte leads to a significant
increase of the surface pressure \cite{Galla10}. The corresponding Pearson correlation coefficient is given by $r=0.93$ which validates a linear dependence between osmolyte concentration and surface pressure.
Although we have studied bilayers, the evident properties of Fig.~\ref{fig5} are in good agreement to the experimental findings in Refs.~\cite{Smiatek12,Galla10,Galla11} for DPPC monolayers.
The direct connection between the surface pressure and the surface tension $\gamma$ in regards to Eqn.~\ref{eq:gamma} also allows to validate an identical behavior as it has been observed for trehalose-DPPC mixtures \cite{Pastor05}.
In regards to the presence of unfavorable environmental conditions for extremophilic organisms, the observed 
effect is advantageous due to the fact that it leads to a fluidization of lipid membranes. It has been discussed in Ref.~\cite{Driller08} that the enhanced flexibility of membranes facilitates cell repair and 
signal transport.\\
As an additional property, we have calculated the solvent accessible surface area (SASA). It can be assumed that a less rigid bilayer coincides with 
an increasing solvent accessible surface area. The results are shown in Fig.~\ref{fig6}. It can be seen that a small increase of the total SASA $\Sigma_{tot}$ for higher hydroxyectoine concentrations is evident. 
Although the overall amount
of this increase compared to the total area is small ($\approx 1.5$ nm$^2$ compared to $43$ nm$^{2}$), the general behavior follows a monotonous increase with a Pearson correlation coefficient of $r=0.97$. 
We have also calculated the ratio of the hydrophilic SASA $\sigma_{HL}$, where only the polar regions of the molecules are taken into account to
the total SASA $\Sigma_{tot}$. The results are shown in the inset of Fig.~\ref{fig6}. A linear increase for this quantity can be also identified. 
The observed behavior can be brought into agreement with the increased surface pressure as shown in Fig.~\ref{fig5}, 
due to the fact that the hydrophilic regions form the interface with the hydroxyectoine solution. We therefore propose that this effect accounts for an increased fluidization of the lipid 
bilayer in agreement to recent experiments \cite{Galla11}.\\
In contrast to the proposed decreased solvent accessible surface for polar protein surfaces in presence of kosmotropes \cite{Collins04}, we have observed the opposite trend for DPPC bilayers.
The difference between molecular complex surfaces like membranes with a more flexible surface area compared to a single molecular surface may be responsible for this observation.\\   
Finally we have calculated the electrostatic potential for the DPPC lipid bilayer in presence of varying hydroxyectoine concentrations. 
The results are presented in Fig.~\ref{fig7}. It can be seen that for higher hydroxyectoine concentrations an increase of the DPPC electrostatic potential can be additionally observed. The general trend is obvious although
a deviation for the concentration $c=0.111$ can be identified which can be related to a slight asymmetric potential distribution compared to the other concentrations. However, these findings are in agreement to
a recent publication \cite{Cebers08}, where it has been found that electrostatic contributions play a significant role for the internal organization of lipid bilayers. It can be assumed that the strong 
zwitterionic charges of
hydroxyectoine interact with the nitrogen atoms and the phosphate groups in DPPC. Due to the fact that the electrostatic potential has been calculated by integrating 
over the charges within a slice, we propose that the accumulation of charged groups at the interface accounts for this observation. This reason has been also discussed in a recent publication \cite{Chachisvilis11},
where the strong dependence of the electrostatic potential on the local ordering and the packing fraction has been pointed out. 
Although the authors have remarked, that the molecular origin for the varying potential remains controversial,
a strong dependence between local ordering of the lipid molecules and the resulting electrostatic potential has been proposed.   
We are therefore confident that the observed behavior validates the ordering effect of lipid bilayers in presence of low hydroxyectoine concentrations.
Hence, it has to be stated that only a combination of many effects may explain the observed characteristics as it has been also proposed in Ref.~\cite{Hunenberger10}.
\section{Summary and conclusion}
Molecular Dynamics simulations of DPPC lipid bilayers in presence of hydroxyectoine have been performed. We have analyzed the properties of the aqueous solution  
for low concentrations of hydroxyectoine. Our results have clearly revealed that the presence of hydroxyectoine results in a strengthening of the water hydrogen bond network due to increased forward life 
times. Due to the fact that the life times are related to the activation free energy barriers, we propose that the water hydrogen bond network is strengthened in presence of hydroxyectoine. Although we have simulated 
very low concentrations of hydroxyectoine, all observed effects are clearly visible. Thus, we were able to show that specific cosolute-water-solute effects can be even observed at low physiological concentrations
in agreement to recent experimental findings. Furthermore we have shown that the presence of hydroxyectoine leads to a energetic decrease of the hydrogen bond network between DPPC and water molecules in terms 
of slightly decreased forward life times. These findings are in good agreement to previous literature results for kosmotropic properties in solution. Summarizing the results for the co-solute-solvent interactions, 
it can be concluded that hydroxyectoine can be interpreted as a typical kosmotropic osmolyte.\\
As a further characteristic property, it has been often proposed that kosmotropic cosolutes are preferentially excluded from polar surfaces \cite{Collins04}. The preferential binding coefficient in our simulations 
has been calculated by the Kirkwood-Buff theory and indicates a preferential exclusion behavior for hydroxyectoine in presence of DPPC bilayers. Thus, our findings are in good agreement to recent theories. 
The distance between the hydrophilic head groups of DPPC and the carboxy group of hydroxyectoine has been determined to fluctuate around 0.5 nm. 
This value is in agreement to recent conclusions \cite{Smiatek12} and 
roughly corresponds to the position of the second hydration shell.\\ 
In regards to the results for the DPPC lipid bilayer, we have indicated an increase of the surface pressure for higher hydroxyectoine concentrations. These results are also in good agreement to recent experimental findings 
for DPPC monolayers. 
We propose that the increase of the surface pressure in presence of hydroxyectoine which has been observed in our simulations is responsible for the
experimentally observed broadening of the LE/LC phase transition in monolayers. The slight swelling of the DPPC bilayer has been also identified by an increased solvent accessible surface area. 
In terms of electrostatic interactions, we have observed that an increased electrostatic potential can be estimated for higher hydroxyectoine concentrations. The molecular origin of this effect is not clear, but 
it was proposed that the local ordering of the DPPC molecules contributes a significant amount to this observation \cite{Chachisvilis11}.
It can be therefore assumed that further contributions may also contribute significantly to the fluidization of membranes in addition to pure hydration effects.\\
In summary, we have shown that the usage of small concentrations of co-solutes in computer simulations will lead to observable effects in agreement to recent experiments. The presence of hydroxyectoine
as a typical kosmotropic co-solute results in an increased surface pressure for DPPC bilayers which accounts for the experimentally observed broadening of the LE/LC phase transition. 
\section{Acknowledgments}
The authors thank Davit Hakobyan and Oliver Rubner for enlightening discussions and helpful remarks.
Financial support by the Deutsche Forschungsgemeinschaft (DFG) through the SFB
858 and the transregional collaborative research center
TRR 61 is gratefully acknowledged.

\end{document}